\documentclass[doublecol]{epl2}
\usepackage{graphicx}

\newcommand\beq{\begin{equation}}
\newcommand\eeq{\end{equation}}
\newcommand\beqa{\begin{eqnarray}}
\newcommand\eeqa{\end{eqnarray}}

\newcommand\e{\epsilon}

\newcommand{\comment}[1]{}

\newcommand\bk{\mathbf{k}}

\title{Long-wavelength fluctuations lead to a model of the  glass crossover}

\author{Tommaso Rizzo \inst{1,2}}

\institute{
\inst{1} Dip. Fisica,
Universit\`a "Sapienza", Piazzale A. Moro 2, I-00185, Rome, Italy \\
\inst{2} IPCF-CNR, UOS Rome, Universit\`a "Sapienza", PIazzale A. Moro 2,
I-00185, Rome, Italy}


\abstract{
The effect of long wavelength fluctuations on the Mode-Coupling-Theory (MCT) dynamical singularity at $T_c$ in the $\beta$ regime is studied by means of the standard field-theoretical procedure for a genuine second-order phase transition. The resulting perturbative loop expansion can be resummed leading to
an extension of the MCT equation for the critical correlator with local random fluctuations of the separation parameter. 
The corresponding model explains both qualitatively and quantitatively why the MCT dynamical singularity is transformed into a crossover from relaxational to activated dynamics. Dynamical Heterogeneities emerge naturally as the ergodicity-restoring mechanism instead of {\it ad hoc} hopping processes.}

\begin{document}

\maketitle

Mode-Coupling-Theory (MCT) provides a rather accurate description of the early stages of the dynamical slowing down in super-cooled glass-forming liquids \cite{Gotze09}.
It makes many predictions in agreement with experiments \cite{Gotze99}, both qualitatively, {\it e.g.} the the stretching of the time correlators relaxation in a two-step fashion, and quantitatively {\it e.g.} the ergodicity-breaking parameter. 
The behavior of the critical correlators and the dynamical exponents in the $\alpha$ and $\beta$ regimes according to the theory \cite{Gotze85} reproduces also quite well the numerical data \cite{Kob97-98,Sciortino01,Weysser10}.

The main problem of the theory is that it predicts at a temperature $T_c$ a divergence of the $\alpha$ relaxation time $\tau_\alpha$ that is not observed nor in experiments neither in numerical simulations.
Nevertheless early numerical studies \cite{Hansen} suggested that $T_c$ marked instead a dynamical crossover from relaxational to  activated dynamics and this is currently the widespread interpretation among those who believe that MCT has not to be dismissed altogether. In order to amend this problem it has been proposed \cite{Gotze87} to add to the theory  some sort of hopping mechanism able to restore ergodicity through activated dynamics. This proposal has been further examined in connection with experimental data \cite{Gotze99,Gotze00} and also alternative hopping models have been proposed over the years \cite{Mayer06,Greenall07,Bhattacharyya08,Chong08}. In spite of these efforts the common belief is that a definite convincing solution to the avoided singularity problem, if any, is still to be found. 
In particular there are two main drawbacks in the aforementioned hopping models. The first one is that hopping is essentially  an {\it ad hoc} assumption superimposed on MCT thereby changing its status from a first-principle theory to a phenomenological one. The second drawback is that while MCT successfully predicts stretched exponential relaxation above $T_c$, hopping dynamics  leads eventually to simple exponential relaxation at and below $T_c$ in contrast with experiments. 

A satisfactory qualitative and quantitative characterization of the dynamical crossover at $T_c$ is highly desirable. We note that some authors \cite{Mallamace11} have suggested that a crossover temperature, identified with $T_c$, is the sole relevant temperature for many glassy materials in contrast to approaches that advocate the presence of an ideal phase transition below the glass transition temperature $T_g$ leading to Vogel-Fulcher-Tammann (VFT) scaling \cite{Adams65,Stillinger88,Kirkpatrick87d}. Besides the fact that the validity of VFT scaling is itself the object of an ongoing debate (see {\it e.g.} \cite{Elmatad09,Zhao13}), a quantitative description of the $T_c$ crossover could tell if below $T_c$ simple Arrhenius relaxation explains the data or a non-trivial mechanism (as proposed {\it e.g. } in Random-First-Order-Theory \cite{Kirkpatrick87d}) is definitively needed.

In this work I discuss the application to MCT of the standard field-theoretical procedure to study the effect of long-wavelength fluctuations on a genuine second-order phase transition \cite{Zinn02}. 
In this kind of analysis one introduces a (dynamical) field-theory
with the structure of the original theory as an effective Landau theory valid at a coarse grained scale (in space and time).
The field theory is then studied by means of a perturbative loop expansion and the resulting series can be recast into a series expansion for the evaluation of the critical exponents, although it is typically limited to a few orders in perturbation theory.
The main result presented in this letter is that instead in the case of MCT one has full control of the loop expansion and can show that it   is equivalent at all orders to a stochastic dynamical model defined by eqs. (\ref{phimediodyn}) and (\ref{stochglass}), called stochastic $\beta$-relaxation (SBR) equations in the following.
The definition of the SBR model is particularly simple: it is {\it an extension of the standard MCT equation for the critical correlator with local random fluctuations of the separation parameter}. 
As I will discuss in the following the model itself does not display any dynamical singularity, $T_c$ is naturally turned into a crossover temperature and a qualitative and quantitative characterization of the $\beta$-time correlator at and below $T_c$ can be obtained.

SBR, as a  model of the glass crossover, has several interesting features. First of all ergodicity is restored not through some obscure hopping mechanism but through fluctuations that lead to localized regions relaxing faster than the rest of the system even below $T_c$. This clearly suggest the presence of strong Dynamical Heterogeneities (DH) which have been established as a key feature of glassy dynamics in the last twenty years \cite{Kob97,Ediger00}.  
The model does not suffer from the aforementioned drawbacks: 1) the ergodicity restoring mechanism is not an {\it ad hoc} assumption but emerges as a non-perturbative effect of an otherwise standard field-theoretical computation, 2) it implies stretched-exponential relaxation below $T_c$ with the same dynamical exponents valid above.
This last feature offers the important possibility of making predictions in the $\alpha$ regime as well.
It is to be expected that this model will extend the range of temperatures where theory can describe experimental observations both qualitatively and quantitatively. Indeed the parameters of the model can be computed through straightforward extensions of standard MCT methods and also connected to physical observables.  The model deals essentially with the $\beta$-regime while more work is needed in order to fully characterize the $\alpha$-regime. Nevertheless this result suggests that a complete understanding of the glass crossover is now at hand. 

The central quantity of MCT is the normalized autocorrelation function of density fluctuations at given wave-vector ${\mathbf k}$
\beq
\Phi(k,t)\equiv \langle \delta\rho^*({\mathbf k},t)\delta\rho({\mathbf k},0)\rangle/S(k)
\eeq
where $S(k)\equiv\langle |\delta\rho(k,0)|^2\rangle$ is the static structure factor.
Below the critical temperature $T_{c}$ MCT predicts a glassy phase: the long-time limit of the correlator (which vanishes in the liquid phase) remains finite: $\lim{t \rightarrow \infty} \, \Phi(\bk,t)=f(k) \neq 0$.

For temperatures near the critical temperature one identifies the $\beta$-regime corresponding to time-scales $\tau_\beta$ when the correlator is almost equal to $f(q)$. In the liquid phase ($T>T_c$) this regime is followed by the $\alpha$-regime during which the correlator decays from $f(k)$ to zero. In the $\beta$-regime the time-dependence of the correlator is controlled by the following scaling law \cite{Gotze85}:
\beq
\Phi(k,t)=f(k)+|\tau|^{1/2} g_{\pm}(t/\tau_\beta) \, \xi_c^R(k)
\label{scalfor}
\eeq
where $\tau$ is a linear function of $T_c-T$, {\it i.e.} it is negative in the liquid phase ($g_-(s)$ has to be used) and positive in the glassy phase ($g_+(s)$ has to be used).
The function $g_{\pm}(s)$ obeys the following scale-invariant quadratic equation:
\beq
\pm 1=g_{\pm}^2(s)\left(1-\lambda\right) +\int_0^s (g_{\pm}(s-s')-g_{\pm}(s))\dot{g}_{\pm}(s')ds'
\label{SVDYN2}
\eeq
For small values of $s$ both the functions $g_{\pm}(s)$ diverge as $1/s^a$, while for large values of $s$ $g_+(s)$ goes to a constant while $g_-(s)$ diverges as $-s^b$ where the exponents $a$ and $b$ are determined by the so-called parameter exponent $\lambda$ according to:
\beq
\lambda={\Gamma^2(1-a)\over\Gamma(1-2a)}={\Gamma^2(1+b)\over\Gamma(1+2b)}
\label{lambdaMCT}
\eeq
The parameter exponent $\lambda$ controls also the time scale of the $\beta$ regime that diverges with $\tau$ from both sides as $\tau_\beta \propto |\tau|^{-1/(2\,a)}$ with an unknown model-dependent factor.  By using matching argument one can also argue that the time-scale of the $\alpha$ regime increases as $\tau_\alpha \propto |\tau|^{-\gamma}$ with $\gamma=1/(2 a)+1/(2 b)$.
Note that the although the density-density correlator $\Phi(k,t)$ depends on the momentum $k$, the behavior near $T_c$ is controlled solely by the critical mode $\xi_c^R(k)$, meaning that the actual critical quantity is a single scalar $\phi(x,t)$, {\it i.e.} the component of $\Phi(k,t)-f(k)$ along the critical mode.
We take $\phi(x,t)$ as the (Landau) order parameter of the problem and
we want to develop a dynamical field-theory in order to study its fluctuations on arbitrarily large space scales because that are not taken into account in the original treatment.  The first candidate field-theory to study critical behavior  at $T_c$ is actually a static field theory. This should not be a surprise, after all if $T_c$ marked a true glass transition it should be possible to characterize the system below $T_c$ with a static theory. Less trivial is the fact that the order parameter of the theory is a replicated version of the correlator $\phi_{ab}(x)$. The theory itself is the  following cubic Replica-Symmetric (RS) field theory with $n=1$ replicas:
\beqa
{\mathcal L} & = &{1 \over 2}\int dx \left[- \tau \sum_{ab}\phi_{ab}+{1 \over 2} \sum_{ab} (\nabla
\phi_{ab})^2+ \right.
\nonumber
\\
&+&m_2\sum_{abc}
\phi_{ab}\phi_{ac}+ m_3\sum_{abcd}\phi_{ab}\phi_{cd} +
\nonumber
\\
& - &\left.{1 \over
  6}w_1 \sum_{abc}\phi_{ab}\phi_{bc}\phi_{ca}-{1 \over 6}w_2
\sum_{ab}\phi_{ab}^3\right]
\label{T3}
\eeqa
This theory arises naturally in the context of the so-called one-step-Replica-Symmetry-Breaking (1RSB) Spin-Glass (SG) models and its relevance for structural glasses is motivated by the discovery that these SG systems are controlled by the very same MCT equations  (\ref{SVDYN2}) and (\ref{lambdaMCT}) \cite{Kirkpatrick87c,Crisanti93}. We refer the reader to \cite{Franz12} for an extensive presentation of the above replicated action in the context of structural glasses and to \cite{Szamel10,Rizzo13} for a discussion of the close relationship between static replica methods and MCT. 

The above static action makes sense only in the glassy phase $\tau>0$ where it can be extremized by the a RS field constant in space $\phi_{ab}(x)=\phi$ given by the solution of the equation of state:
\beq
\tau =\left(w_1-w_2\right)\phi^2 \ .
\eeq
One can then study systematically the loop expansion around the mean-field solution. Quite surprisingly it has been recently discovered \cite{Franz11b} that the loop expansion is equivalent at all orders to a stochastic equation.
For instance the thermal average of the order parameter $\phi$ in the glassy phase is given by:
\beq
\langle \phi(x) \rangle = [\phi_{\tau+\epsilon}(x)]_\epsilon
\label{phimedio}
\eeq
where the square brackets mean average with respect to a Gaussian distributed random field $\epsilon(x)$ with variance
\beq
[\epsilon(x)\epsilon(y)]=-4 (m_2+m_3) \delta (x-y) 
\label{fluca}
\eeq
and $\phi_{\tau+\epsilon}(x)$ is the solution of the following quadratic equation:
\beq
\tau + \epsilon(x) =-\nabla^2 \, \phi +\left(w_1-w_2\right)\phi^2(x)
\eeq
This result poses a serious problem: in the thermodynamic limit there will be always fluctuations of the temperature that drive portions of the system in the liquid phase, as a consequence the stochastic equation has no real solution and the whole static construction is ill-defined. However we are happy with this because it implies that the glassy phase is unstable in agreement with all expectations.

In order to obtain a characterization of the dynamical crossover one has to put dynamics back into the problem. I have done this in such a way to keep the computation as close as possible to the static replica computation by means of a  a superfield description of the dynamics. In that context it has been argued \cite{Parisi13} that the dynamical field theory of the super-field correlator has the {\it same} structure of the replicated field theory (\ref{T3}) with the {\it same} coupling constants. Most importantly when the equations of state for this critical super-field correlator are translated into the those of the standard correlator one finds \cite{Parisi13} that they have precisely the structure of the MCT critical equation (\ref{SVDYN2}):  
\beq
\tau =\left(w_1-{w_2}\right)\phi^2(t)+w_1\int_0^t (\phi
(t-t')-\phi (t))\dot{\phi}(t') dt'
\label{SPdyn}
\eeq
from which one identifies \cite{calta1,Parisi13}: 
\beq
\lambda={w_2 \over w_1} \ .
\eeq 
To lighten the notation in the following I will fix the normalization of the critical mode $\xi_c^R(k)$ such that $w_1=1$ without loss of generality.
I have computed (details elsewhere) perturbative loop corrections to the {\it dynamical} field theory with the structure (\ref{T3}) around the dynamical solution (\ref{SPdyn}) for the critical correlator in the $\beta$-regime. The first step is the computation of the  the scaling form equivalent to (\ref{SVDYN2}) for the bare propagator that is essentially the four-point susceptibility studied intensively in  recent times \cite{Berthier07}.  Then one has to determine the rules to evaluate all possible diagrams from which a mapping to a stochastic equation can be shown at all orders following Parisi and Sourlas
\cite{Parisi79}. 
In the end the solution is still of the form (\ref{phimedio}): 
\beq
\langle \phi(x,t) \rangle = [\phi_{\tau+\epsilon}(x,t)]_\epsilon
\label{phimediodyn}
\eeq
with the difference that $\phi_{\tau+\epsilon}(x,t)$ is now the solution of the following {\it dynamical} equation:
\beqa
\tau  +  \epsilon(x)    =    -\nabla^2 \, \phi(x,t)+\left(1-\lambda\right)\phi^2(x,t)+
\nonumber
\\
 +   \int_0^t (\phi
(x,t-t')-\phi (x,t)){d {\phi} \over dt'}(x,t') dt'
\label{stochglass}
\eeqa
The above equations defines the SBR model for the behavior of the $\beta$-correlator in the region of the avoided dynamical singularity. The correlator is the average over different solutions of eq. (\ref{stochglass}) which is an extension of the MCT critical equation with a separation parameter with local Gaussian fluctuation given by (\ref{fluca}).
In order to understand the model one must bear in mind that it shares many of the essential features of eq. (\ref{SVDYN2}), in particular: i) the above equation is time scale-invariant and for all $x$ the field $\phi(x,t)$ diverges at small times as $1/t^a$; the actual constant is the same for all $x$ but it is fixed by the microscopic details of the system \footnote{One can follow the convention  \cite{Gotze09} that  $\lim_{t \rightarrow 0} \phi(x,t) t^a=1$}. ii) the equation is only valid provided $\phi(x,t)$ is small, {\it i.e.} when the correlator is near the non-ergodicity parameter corresponding to the plateau of the correlator; however in the liquid solutions $\phi(x,t)$ goes to minus infinity at large times, this corresponds to the correlator entering the $\alpha$ regime where the model is no longer valid. 

Many features of the model can be understood considering a simplified version with no space dependence \footnote{The resulting model is of direct relevance for the class of mean-field discontinuous spin-glass models defined on random-lattices}. In this case the solutions of eq. (\ref{stochglass}) are given by $g_{\pm}(s)$ with the appropriate factor and scale determined by $\tau+\epsilon$. One easily sees that well below $T_c$ ($\tau \gg 1$) the majority of the solutions will never decay, however there will always be also liquid solutions decaying from the plateau as $-t^b$. At large enough times these liquid solutions will eventually dominate but they are exponentially rare because they requires $\tau+\epsilon<0$. Using the standard matching argument between the $\beta$ and $\alpha$ regimes \cite{Gotze85} one can obtain an estimate of $\tau_\alpha$ as a function of the $\tau$. An explicit computation shows that the divergence of $\tau_\alpha$ at $T_c$ ($\tau=0$) is replaced by a crossover from power-law increase at high temperatures ($\tau<0$) to exponential increase at low temperatures ($\tau>0$):
\beq 
\tau_\alpha \propto   \left\{ \begin{array}{ll}
 |\tau|^{-\gamma} & \mbox{for $\tau \rightarrow  -\infty$};\\
        e^{\tau^2 \over 2 b} \tau^{1/b+\gamma} & \mbox{for $\tau \rightarrow  \infty$}.\end{array} \right. 
\eeq 

A detailed study of the SBR equations is left for future work. Such a study will yield crossover functions for a number of quantities that are otherwise singular at $T_c$ in MCT, including for instance $\tau_\alpha$, $\tau_\beta$ and the Debye-Waller factor \cite{Gotze99}. 

The solution of the full space-dependent model should also provide a quantitative characterization of DH.
Qualitatively we expect that in the region of high temperatures (large negative values of $\tau$) $\tau$ dominate both over the random field and on the gradient term that can be both treated perturbatively. This is the region where the standard MCT scalings apply. The dynamical correlation length will also increase with the mean-field exponent $\xi \propto |\tau|^{-1/4}$ in agreement with previous results \cite{Biroli06}.
This state of things changes when the corrections start to be relevant. Approaching $\tau=0$ the transition is avoided and $\tau_\alpha$, $\tau_\beta$ and $\xi$ remain finite.   In this regime one should observe strong DH, because small fluctuations of the separation parameter induce large fluctuations of the local relaxation times. Lowering the temperature well below $T_c$, $\tau$ is again large an positive and we enter the activated regime. In this regime the typical solution appears to be frozen; its eventual relaxation is due to rare regions where $\tau+\e(x)$ is negative (corresponding to the liquid) because of fluctuations with exponentially small probability.
On the other hand scaling suggests that the actual size of these regions {\it decreases} in the deep activated regime ($\tau \gg 1$) and one may ask if there is a connection to  observations of a non-monotonous behavior across $T_c$ of a properly defined dynamical correlation length \cite{Kob12}. Clearly the size of the liquid regions cannot decrease beyond the microscopic scale and below a certain temperature the continuous stochastic glassy equation must be abandoned. What happens then requires a different analysis, standard Arrhenius behavior being a possibility.

It is natural to identify each solution of the model with a realization of the dynamics starting from  a given initial configuration of the system. Consistently in finite dimension all equilibrium initial configurations should be equivalent because distant regions are uncorrelated and we expect that each solution has essentially the same behavior in the thermodynamic limit (in particular the correlator will always eventually decay to minus infinity). We stress however that local random fluctuations that determine the position of DH in eq. (\ref{stochglass}) have no time dependence, this can be true in the $\beta$ regime but not in the $\alpha$ regime because ergodicity must be restored after large times. Therefore a complete description of the dynamics near $T_c$ requires a characterization of the $\alpha$ regime where eq. (\ref{stochglass}) is no longer valid and one should be careful in extracting information solely by means of matching arguments.

The quantitative predictive power of the SBR equations is open to tests against experiments and numerical simulations. In doing these kind of comparisons one should bear in mind that  the model is the result of the resummation of the most divergent corrections to MCT, but since the theory in the end turns out not be critical there is no well-defined scaling regime in which the model holds exactly. Furthermore the range of temperatures (how far below $T_c$?) and of times over which the model provides an accurate description of the dynamics are expected to be system-dependent.
I have chosen not to present here any comparison with actual data because
it could have been misleading: this is not just another phenomenological theory of the glass crossover that should be be judged on the basis of its ability to fit experimental data in a the largest possible parameter window. Actually, as I explained before, the SBR equations are certainly not supposed to give an accurate quantitative description far from $T_c$. The key point here is that without making any {\it ad hoc} assumption on activated process one performs a complex computation and ends up with a model where there is no dynamical arrest. This is far more important than the actual ability of eqs. (\ref{stochglass}) to describe quantitatively the crossover with high accuracy. Various neglected corrections terms can be now added to the theory in order to improve its quantitative predictions but none of these corrections alone would have been able to turn the transition into a crossover.

The theory does not make any quantitative prediction on the (pseudo) critical temperature, the parameter exponent $\lambda$, the variance of the separation parameter and so on: all this information must be put into it at the beginning specifying the values of the coupling constants in expression (\ref{T3}).
Therefore it is not to be expected to improve the estimate of the critical temperature, which is known to be inaccurate in MCT. I note that a reasonable explanation of this problem  is that since the critical temperature corresponds to the vanishing of the coefficient of the linear term in the action (\ref{T3}) it is more sensible to approximations than quantities ({\it e.g.} $\lambda$) that are controlled by the constants that instead remain finite at $T_c$.  Besides $\lambda$ and $\tau$,  the variance of the random temperature can be obtained by reading the coefficients $m_2$ and $m_3$ from the static replica formulation of MCT \cite{Szamel10,Rizzo13}.
Less trivial is the computation of the coefficient of the Laplacian that should be estimated by means of the inhomogeneous MCT extension discussed in \cite{Biroli06}. Alternatively one could use the (less accurate) estimates for the coupling constants of (\ref{T3}) from the Hypernetted-Chain-Approximation \cite{Franz12}.  Lastly we recall that in principle the coupling constants can be also extracted directly from measurements of appropriate non-linear susceptibilities \cite{Parisi13}.

Concluding, we have shown that a standard field-theoretical treatment of the the dynamical arrest transition of MCT unveils its avoided nature leading to an intuitive model of the glass crossover. The simplicity of the final result could be misleading: one should not forget that it is the result of a complex computation and should not  confuse the conventional long wave-length fluctuations of the order parameter (the starting point of the computation) with the fluctuations of the separation parameter of MCT that are a completely unexpected outcome.

\begin{acknowledgments}
I thank G. Parisi, L. Leuzzi, G. Biroli, M. Fuchs, T. Voigtmann and G. Szamel for discussions.
\end{acknowledgments}

\end{document}